\documentclass[11pt]{article}
\usepackage{epsfig}
\usepackage{amssymb}
\usepackage[]{times}

\makeatletter
\@addtoreset{equation}{section}

\makeatother
\newcommand{\be}{\begin{equation}}
\newcommand{\ee}{\end{equation}}
\newcommand{\bea}{\begin{eqnarray}}
\newcommand{\eea}{\end{eqnarray}}

\begin{document}
\title{Multiple Liquid Bridges with Non-Smooth Interfaces}
\author{Leonid G. Fel$^1$,
Boris Y. Rubinstein$^2$ and Vadim Ratner$^3$\\ \\
$^1$Department of Civil and Enviromental Engineering, \\
Technion -- Israel Institute of Technology, Haifa, 32000, Israel,\\
$^2$Stowers Institute for Medical Research, 1000 E 50th St, Kansas City,
MO 64110, USA,\\
$^3$Department of Computer Science, Stony Brook University,\\ Stony Brook,
NY 11794-2424, USA}
\date{}
\maketitle
\begin{abstract}
We consider a coexistence of two axisymmetric liquid bridges $LB_i$ and $LB_m$
of two immiscible liquids {\em i} and {\em m} which are immersed in a third
liquid (or gas) {\em e} and trapped between two smooth solid bodies with
axisymmetric surfaces $S_1,S_2$ and free contact lines. Evolution of liquid 
bridges allows two different configurations of $LB_i$ and $LB_m$ with multiple
(five or three) interfaces of non-smooth shape. We formulate a variational 
problem with volume constraints and present its governing equations supplemented
by boundary conditions. We find a universal relationship between curvature of 
the interfaces and discuss the Young relation at the singular curve where all 
liquids meet together.
{\bf PACS 2006:} Primary -- 76B45, Secondary -- 53A10
\end{abstract}
\section{Introduction}\label{c1}
\begin{figure}[h!]\begin{center}\begin{tabular}{cc}
\psfig{figure=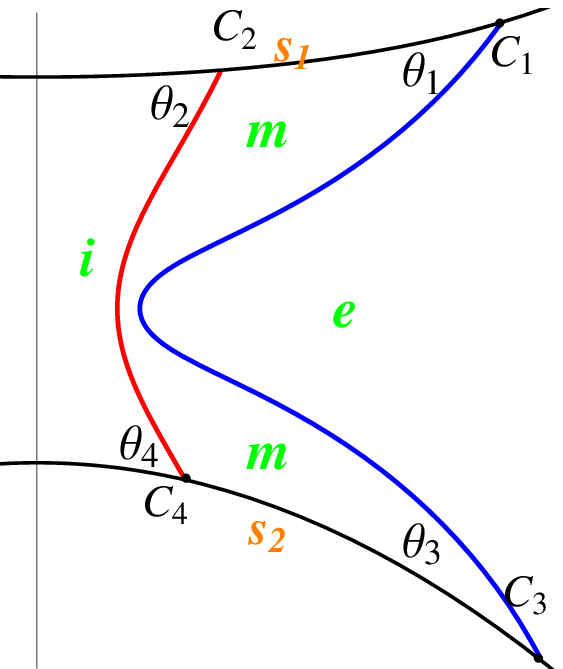,height=5cm}
\hspace{1cm}&\hspace{1cm}
\psfig{figure=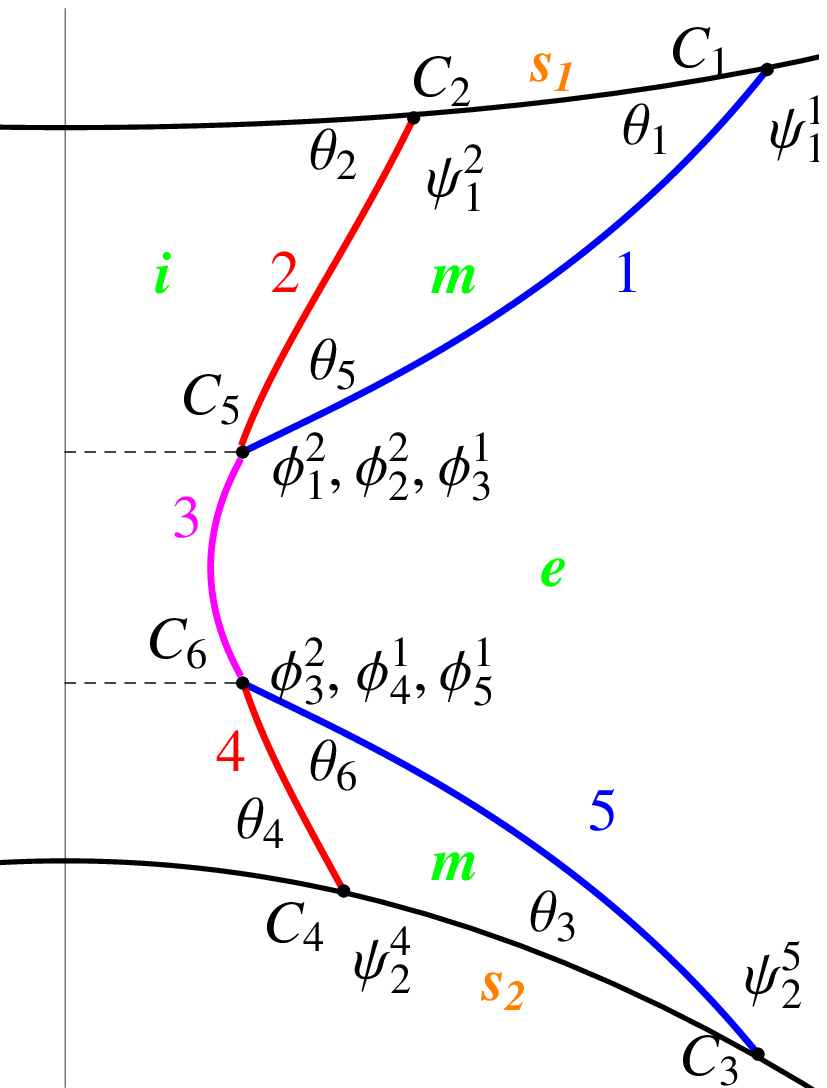,height=5cm}\\
(a)&(b)\\
\end{tabular}\end{center}\vspace{-.2cm}
\caption{a) The meridional section of two ${\sf Und}$ interfaces before $LB_m$ 
rupture. b) Five ${\sf Und}$ interfaces of different curvatures for three 
immiscible liquids trapped between two smooth solid bodies with free BC. The 
endpoints $C_1,C_2,C_3,C_4$ have one degree of freedom: the upper and lower 
endpoints are running along $S_1$ and $S_2$, respectively. The endpoints $C_5,
C_6$ have two degrees of freedom and are located on two singular curves $L_1,
L_2$, respectively, which are passing transversely to the plane of Figure.}
\label{f1}
\end{figure}

\begin{figure}[h!]\begin{center}\begin{tabular}{cc}
\psfig{figure=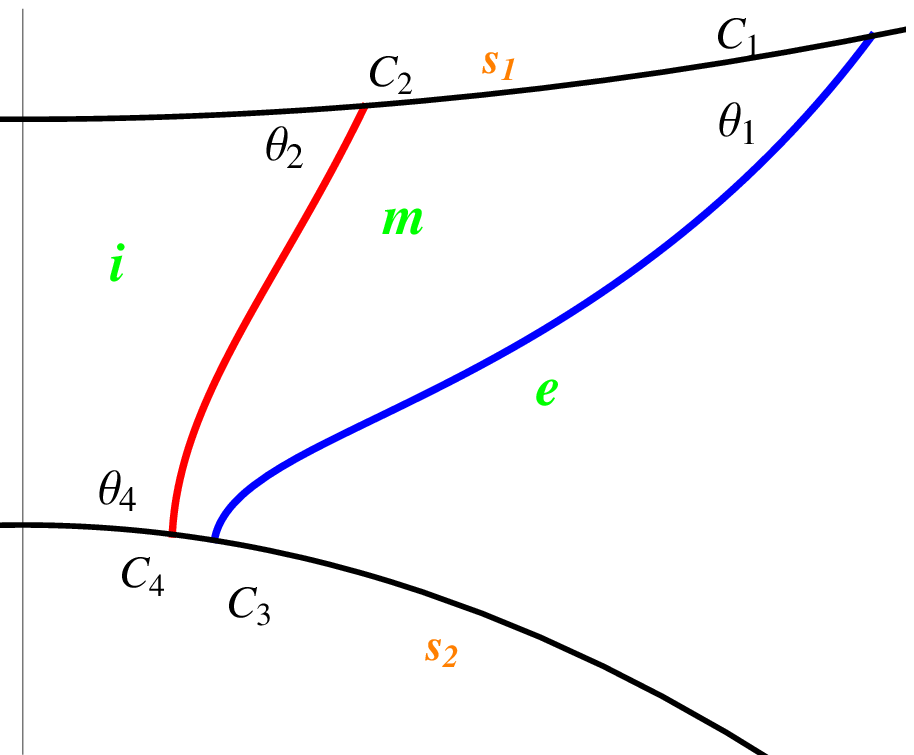,height=5cm}
\hspace{1cm}&\hspace{1cm}
\psfig{figure=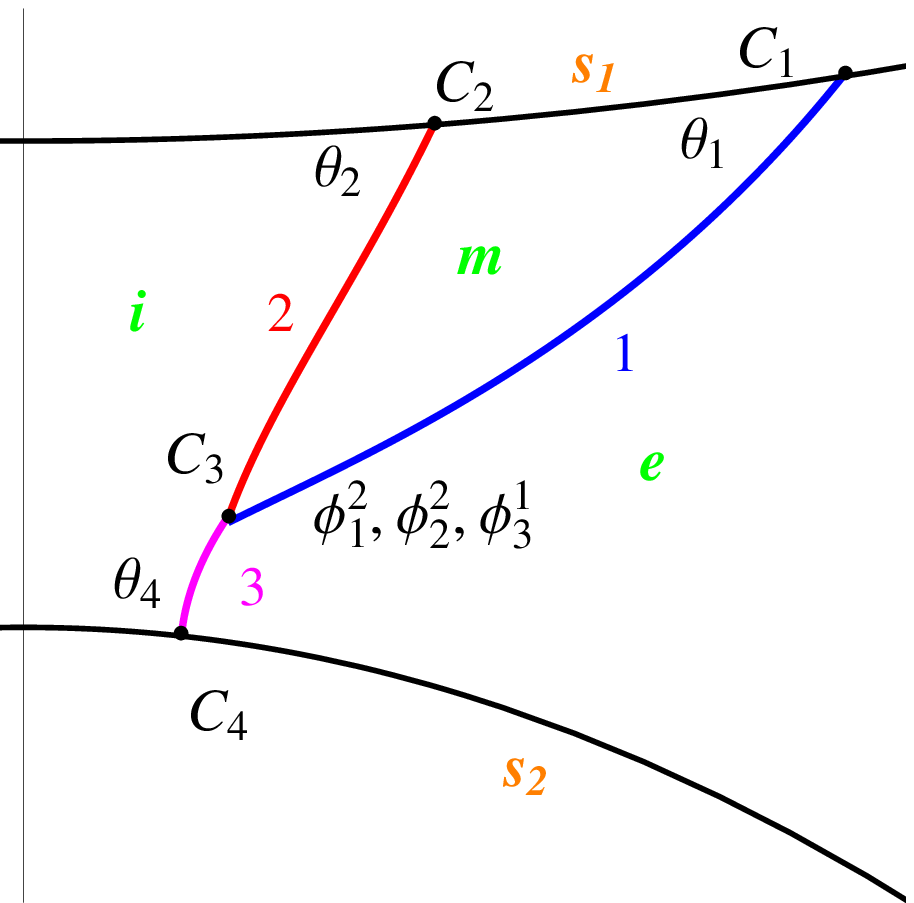,height=5cm}\\
(a)&(b)\\
\end{tabular}\end{center}\vspace{-.2cm}
\caption{a) Two ${\sf Und}$ interfaces before $LB_m$ rupture. b) Three ${\sf 
Und}$ interfaces of different curvatures for three immiscible liquids trapped 
between two smooth solid bodies with free BCs. The endpoints $C_1,C_2,C_4$ have 
one degree of freedom while $C_3$ has two degrees and is located on a singular 
curve $L$ which is passing transversely to the plane of Figure. .}\label{f2}
\end{figure}

Consider an evolution of two liquid bridges $LB_i$ and $LB_m$ of immiscible
liquids, {\em i} (inner) and {\em m} (intermediate), trapped between two
axisymmetric smooth solid bodies with surfaces $S_1,S_2$ in such a way that
$LB_i$ is immersed into $LB_m$ and the latter is immersed into the {\em e}
(external) liquid (or gas) which occupies the rest of the space between the two
bodies (see Figure \ref{f1}a). When liquid {\em m} begins to evaporate then
$LB_m$ reduces in volume (and width). Depending on the relationships between the
contact angles of both liquids on $S_1$ and $S_2$ there are two scenarios for 
connectivity breakage of the liquid bridge {\em m} between the two solids. The 
first scenario ({\em five interfaces}) occurs when $LB_m$ splits into two parts 
each supported by a different solid (see Figure \ref{f1}b).
The second scenario ({\em three interfaces}) occurs when $LB_m$ is left as a 
whole but has support only on the upper (see Figure \ref{f2}b) or lower solid.

Both scenarios lead to a new phenomenon which has not been discussed in
literature before, namely, an existence of multiple LBs with non-smooth 
interfaces. In contrast to the known LBs with fixed and free contact line (CL), 
here one of CLs appears as a line where three interfaces with different 
curvatures meet together. From a mathematical standpoint this singular curve 
is governed by transversality conditions (in physics they are referred to as the
Young relations), and coincidence conditions, i.e., three interfaces always 
intersect at one single curve. We derive a relationship combining the constant 
mean curvatures of three different interfaces and give the interfaces 
consistency rules for their coexistence. Another important result is the 
vectorial Young relation at the triple point which is located on a singular 
curve.
\section{Variational problem for five interfaces}\label{c2}
Consider a functional $E[r,z]$ of surface energy
\bea
E[r,z]&=&\sum_{j=1}^5\int_{\phi_j^2}^{\phi_j^1}{\sf E}_jd\phi_j+
\int_0^{\psi_1^2}{\sf A}_{s_1}^id\psi_1+
\int_{\psi_1^2}^{\psi_1^1}{\sf A}_{s_1}^md\psi_1+
\int_{\psi_1^1}^{\infty}{\sf A}_{s_1}^ed\psi_1+\nonumber\\
&&\hspace{3cm}\int_0^{\psi_2^4}{\sf A}_{s_2}^i\psi_2+\int_{\psi_2^4}^{\psi_2^5}
{\sf A}_{s_2}^md\psi_2+\int_{\psi_2^5}^{\infty}{\sf A}_{s_2}^e\psi_2,\label{e1}
\eea
\bea
{\sf E}_j=\gamma_jr_j\sqrt{r_j'^2+z_j'^2},\;\;1\leq j\leq 5,\quad {\sf A}_{s_
{\alpha}}^l=\gamma_{s_{\alpha}}^lR_{\alpha}\sqrt{R_{\alpha}'^2+Z_{\alpha}'^2},
\quad \alpha=1,2,\;\;l=i,m,e\label{e2}
\eea
where $r_j'=dr_j/d\phi_j$, $z_j'=dz_j/d\phi_j$, $R_{\alpha}'=dR_{\alpha}/d\psi_
{\alpha}$ and $Z_{\alpha}'=dZ_{\alpha}/d\psi_{\alpha}$. Throughout the paper the
Latin and Greek indices enumerate the interfaces and solid surfaces, 
respectively. The surface tension coefficients $\gamma_1=\gamma_5$, $\gamma_2=
\gamma_4$ and $\gamma_3$ denote tension at the {\em e--m}, {\em m--i} and {\em 
e--i} liquid interfaces, respectively, while $\gamma_{s_{\alpha}}^l$ stand for 
surface tension coefficients at the solid-liquid, {\em $s_{\alpha}$--l}, 
interfaces (see Figure \ref{f1}b).

Two other functionals $V_i[r,z]$ and $V_m[r,z]$ for volumes of liquids $i$ and
$m$ read
\bea
V_m[r,z]=\int_{\phi_1^2}^{\phi_1^1}\!{\sf V}_1d\phi_1-\int_{\phi_2^2}^{\phi
_2^1}\!{\sf V}_2d\phi_2+\int_{\phi_5^2}^{\phi_5^1}\!{\sf V}_5d\phi_5-\int_{
\phi_4^2}^{\phi_4^1}\!{\sf V}_4d\phi_4-\int_{\psi_1^2}^{\psi_1^1}{\sf B}_{s_1}
d\psi_1+\int_{\psi_2^4}^{\psi_2^5}{\sf B}_{s_2}d\psi_2,\nonumber\\
V_i[r,z]=\int_{\phi_2^2}^{\phi_2^1}{\sf V}_2d\phi_2+\int_{\phi_3^2}^{\phi_3^1}
\!\!\!{\sf V}_3d\phi_3+\int_{\phi_4^2}^{\phi_4^1}\!\!\!{\sf V}_4d\phi_4-
\int_0^{\psi_1^2}{\sf B}_{s_1}d\psi_1+\int_0^{\psi_2^4}{\sf B}_{s_2}d\psi_2,
\hspace{2.5cm}\label{e3}
\eea
where
\bea
{\sf V}_j=\frac1{2}z'_jr_j^2,\;\;1\leq j\leq 5,\quad {\sf B}_{s_{\alpha}}=
\frac1{2}Z_{\alpha}'R_{\alpha}^2,\quad\alpha=1,2.\nonumber
\eea

The isoperimetric problem requires to find a set of functions
$\bar{r}_j(\phi_j),\bar{z}_j(\phi_j)$, providing a local minimum of $E[r,z]$
with two constraints $V_i[r,z]=V_i$ and $V_m[r,z]=V_m$ imposed on the volumes
of liquids $i$ and $m$. Consider a composite functional
\bea
W[r,z]=E[r,z]-\lambda_1V_m[r,z]-\lambda_3V_i[r,z],\label{e4}
\eea
with two Lagrange multipliers $\lambda_j$ and represent it in the following
form,
\bea
W[r,z]&=&\sum_{j=1}^5\int_{\phi_j^2}^{\phi_j^1}{\sf F}_jd\phi_1+\int_0^
{\psi_1^2}{\sf G}_1^id\psi_1+\int_{\psi_1^2}^{\psi_1^1}{\sf G}_1^md\psi_1
+\int_{\psi_1^1}^{\infty}{\sf G}_1^ed\psi_1-\nonumber\\
&&\hspace{3cm}\int_0^{\psi_2^4}{\sf G}_2^id\psi_2-\int_{\psi_2^4}^{\psi_2^5}
{\sf G}_2^md\psi_2-\int_{\psi_2^5}^{\infty}{\sf G}_2^ed\psi_2,\label{e5}
\eea
where ${\sf F}_j={\sf F}(r_j,z_j,r_j',z_j')$ and ${\sf G}_{\alpha}^l(R_{\alpha},
Z_{\alpha},R_{\alpha}',Z_{\alpha}')$ are given as follows,
\bea
&&{\sf F}_1={\sf E}_1-\lambda_1{\sf V}_1,\quad {\sf F}_2={\sf E}_2-\lambda_2
{\sf V}_2,\quad {\sf F}_3={\sf E}_3-\lambda_3{\sf V}_3,\quad{\sf F}_4={\sf E}_4
-\lambda_4{\sf V}_4,\quad {\sf F}_5-\lambda_5{\sf V}_5,\hspace{1cm}\nonumber\\
&&{\sf G}_1^i=\lambda_3{\sf B}_{s_1}+{\sf A}_{s_1}^i,\quad{\sf G}_1^m=\lambda_1
{\sf B}_{s_1}+{\sf A}_{s_1}^m,\hspace{1.5cm}\lambda_2=\lambda_3-\lambda_1,\quad
\lambda_5=\lambda_1\quad\lambda_4=\lambda_2,\label{e6}\\
&&{\sf G}_2^i=\lambda_3{\sf B}_{s_2}-{\sf A}_{s_2}^i,\quad{\sf G}_2^m=\lambda_1
{\sf B}_{s_2}-{\sf A}_{s_2}^m,\quad{\sf G}_2^e=-{\sf A}_{s_2}^e,\quad
{\sf G}_1^e={\sf A}_{s_1}^e.\nonumber
\eea

Calculate first variation of $W$ when the functions $\bar{r}_j(\phi_j)$ and
$\bar{z}_j(\phi_j)$ are perturbed by $u_j(\phi_j)$ and $v_j(\phi_j)$,
respectively,
\bea
\delta W\!=\!\sum_{j=1}^5\int_{\phi_j^2}^{\phi_j^1}\!\!\Delta{\sf F}_j\;d\phi
_j\!+\!\left({\sf G}_1^i-{\sf G}_1^m\right)\delta\psi_1^2\!+\!\left({\sf G}_1^m-
{\sf G}_1^e\right)\delta\psi_1^1\!-\!\left({\sf G}_2^i-{\sf G}_2^m\right)
\delta\psi_2^4\!-\!\left({\sf G}_2^m-{\sf G}_2^e\right)\delta\psi_2^5,
\quad\label{e8}\\
{\sf G}_1^i-{\sf G}_1^m={\sf A}_{s_1}^i-{\sf A}_{s_1}^m+\lambda_2{\sf B}_{s_1}
\quad {\sf G}_1^m-{\sf G}_1^e={\sf A}_{s_1}^m-{\sf A}_{s_1}^e+
\lambda_1{\sf B}_{s_1},\hspace{2.7cm}\nonumber\\
{\sf G}_2^i-{\sf G}_2^m={\sf A}_{s_2}^i-{\sf A}_{s_2}^m+\lambda_2{\sf B}_{s_2},
\quad {\sf G}_2^m-{\sf G}_2^e={\sf A}_{s_2}^m-{\sf A}_{s_21}^e+
\lambda_1{\sf B}_{s_2},\hspace{2.3cm}\nonumber\\
\Delta{\sf F}_j=\frac{\partial {\sf F}_j}{\partial r_j}u_j+\frac{\partial
{\sf F}_j}{\partial r_j'}u_j'+\frac{\partial {\sf F}_j}{\partial z_j}v_j+
\frac{\partial {\sf F}_j}{\partial z_j'}v_j'.\hspace{6cm}\label{e9}
\eea
The functions $u_j(\phi_j)$ and $v_j(\phi_j)$ may be derived using a requirement
that the upper free endpoints of the first and second interfaces at Figure  
\ref{f1}b are running along $S_1$ and the lower free endpoints of the forth and 
fifth interfaces - along $S_2$ ,
\bea
&&\bar{r}_j(\phi_j^{\alpha})=R_{\alpha}(\psi_{\alpha}^j),\quad\bar{r}_j(\phi_j^
{\alpha})+u_j(\phi_j^{\alpha})=R_{\alpha}(\psi_{\alpha}^j+\delta\psi_{\alpha}^j)
,\quad u_j(\phi_j^{\alpha})=\frac{dR_{\alpha}}{d\psi_{\alpha}}\delta
\psi_{\alpha}^j,\hspace{1cm}\label{e10}\\
&&\bar{z}_j(\phi_j^{\alpha})=Z_{\alpha}(\psi_{\alpha}^j),\quad\bar{z}_j
(\phi_j^{\alpha})+v_j(\phi_j^{\alpha})=Z_{\alpha}(\psi_{\alpha}^j+\delta\psi_
{\alpha}^j),\quad v_j(\phi_j^{\alpha})=\frac{d Z_{\alpha}}{d\psi_{\alpha}}
\delta\psi_{\alpha}^j,\label{e11}
\eea
where $\alpha=1$ for $j=1,2$ and $\alpha=2$ for $j=4,5$. Substitute (\ref{e9})
into (\ref{e8}) and integrate by parts
\bea
\delta W&=&\sum_{j=1}^5\left[\int_{\phi_j^2}^{\phi_j^1}\left(u_j\frac{\delta
{\sf F}_j}{\delta r_j}+v_j\frac{\delta {\sf F}_j}{\delta z_j}\right)d\phi_j+
\left(u_j\frac{\partial {\sf F}_j}{\partial r_j'}+v_j\frac{\partial {\sf F}_j}
{\partial z'_j}\right)_{\phi_j^2}^{\phi_j^1}\right]+\label{e12}\\
&&\left({\sf G}_1^i-{\sf G}_1^m\right)\delta\psi_1^2+
\left({\sf G}_1^m-{\sf G}_1^e\right)\delta\psi_1^1-
\left({\sf G}_2^i-{\sf G}_2^m\right)\delta\psi_2^4-
\left({\sf G}_2^m-{\sf G}_2^e\right)\delta\psi_2^5,\nonumber
\eea
where $\delta {\sf F}/\delta y_j=\partial {\sf F}/\partial y_j-d/dx\left(
\partial {\sf F}/\partial y_j'\right)$, denotes the variational derivative for
the functional $\int {\sf F}\left(x,y_j,y_j'\right)dx$. Since $u_j(\phi)$ and
$v_j(\phi)$ are independent functions, vanishing of the integral part of
$\delta W$ in (\ref{e10}) gives rise to the Young-Laplace equations (YLE)
\cite{FR2014},
\bea
\frac{\delta {\sf F}_j}{\delta r_j}=0,\quad\frac{\delta {\sf F}_j}{\delta z_j}=0
\quad\rightarrow\quad\frac{z_j'}{r_j}+z_j''r_j'-z_j'r_j''=\frac{\lambda_j}
{\gamma_j},\quad 1\leq j\leq 5.\label{e13}
\eea
Setting the remaining terms (\ref{e12}) to zero gives rise to the four 
transversality relations,
\bea
&&\frac{d R_1}{d\psi_1}\left(\psi_1^1\right)\frac{\partial {\sf F}_1}{\partial
r_1'}\left(\phi_1^1\right)+\frac{d Z_1}{d\psi_1}\left(\psi_1^1\right)\frac{
\partial {\sf F}_1}{\partial z'_1}\left(\phi_1^1\right)+
{\sf G}_1^m\left(\psi_1^1\right)-{\sf G}_1^e\left(\psi_1^1\right)=0\nonumber\\
&&\frac{d R_1}{d\psi_1}\left(\psi_1^2\right)\frac{\partial {\sf F}_2}{\partial
r_2'}\left(\phi_2^1\right)+\frac{d Z_1}{d\psi_1}\left(\psi_1^2\right)\frac{
\partial{\sf F}_2}{\partial z'_2}\left(\phi_2^1\right)+{\sf G}_1^i
\left(\psi_1^2\right)-{\sf G}_1^m\left(\psi_1^2\right)=0,\nonumber\\
&&\frac{d R_2}{d\psi_2}\left(\psi_2^4\right)\frac{\partial {\sf F}_4}{\partial
r_4'}\left(\phi_4^2\right)+\frac{d Z_2}{d\psi_2}\left(\psi_2^4\right)\frac{
\partial {\sf F}_4}{\partial z'_4}\left(\phi_4^2\right)+{\sf G}_2^i\left(
\psi_2^4\right)-{\sf G}_2^m\left(\psi_2^4\right)=0,\label{e14}\\
&&\frac{d R_2}{d\psi_2}\left(\psi_2^5\right)\frac{\partial {\sf F}_5}{\partial
r_5'}\left(\phi_5^2\right)+\frac{d Z_2}{d\psi_2}\left(\psi_2^5\right)\frac{
\partial {\sf F}_5}{\partial z'_5}\left(\phi_5^2\right)+{\sf G}_2^m
\left(\psi_2^5\right)-{\sf G}_2^e\left(\psi_2^5\right)=0,\nonumber
\eea
and one more transversality relation
\bea
&&u_1\left(\phi_1^2\right)\frac{\partial {\sf F}_1}{\partial r_1'}\!+\!
v_1\left(\phi_1^2\right)\frac{\partial {\sf F}_1}{\partial z'_1}\!+\!
u_2\left(\phi_2^2\right)\frac{\partial {\sf F}_2}{\partial r_2'}\!+\!
v_2\left(\phi_2^2\right)\frac{\partial {\sf F}_2}{\partial z'_2}\!-\!
u_3\left(\phi_3^1\right)\frac{\partial {\sf F}_3}{\partial r_3'}\!-\!
v_3\left(\phi_3^1\right)\frac{\partial {\sf F}_3}{\partial z'_3}\!+\!
\hspace{.3cm}\label{e15}\\
&&u_3\left(\phi_3^2\right)\frac{\partial {\sf F}_3}{\partial r_3'}\!+\!
v_3\left(\phi_3^2\right)\frac{\partial {\sf F}_3}  {\partial z'_3}\!-\!
u_4\left(\phi_4^1\right)\frac{\partial {\sf F}_4}{\partial r_4'}\!-\!
v_4\left(\phi_4^1\right)\frac{\partial {\sf F}_4}{\partial z'_4}\!-\!
u_5\left(\phi_5^1\right)\frac{\partial {\sf F}_5}{\partial r_5'}\!-\!
v_5\left(\phi_5^1\right)\frac{\partial {\sf F}_5}{\partial z'_5}=0.\hspace{1cm}
\nonumber
\eea
In case of one liquid bridge $LB_m$ and two immiscible liquids {\em m} and {\em
e} between two smooth solids $S_1,S_2$ the first and forth relations in 
(\ref{e14}) coincide with those derived in \cite{FR2014}, formula (2.15), while
the rest of relations disappear. Regarding condition (\ref{e15}), the
perturbations $u_j\left(\phi_j^k\right)$ and $v_j\left(\phi_j^k\right)$ are
related in such a way that the three disturbed interfaces $1,2,3$ (and 
other three $3,4,5$) always intersect at one point,
\bea
u_1\left(\phi_1^2\right)=u_2\left(\phi_2^2\right)=u_3\left(\phi_3^1\right),\quad
v_1\left(\phi_1^2\right)=v_2\left(\phi_2^2\right)=v_3\left(\phi_3^1\right),
\label{e16}\\
u_3\left(\phi_3^2\right)=u_4\left(\phi_4^1\right)=u_5\left(\phi_5^1\right),\quad
v_3\left(\phi_3^2\right)=v_4\left(\phi_4^1\right)=v_5\left(\phi_5^1\right).
\nonumber
\eea
Combine (\ref{e15}), (\ref{e16}) and use independence of $u_1\left(\phi_1^2
\right)$, $v_1\left(\phi_1^2\right)$, $u_3\left(\phi_3^2\right)$, $v_3\left(
\phi_3^2\right)$ and obtain four relations,
\bea
\frac{\partial {\sf F}_1}{\partial r_1'}\left(\phi_1^2\right)+
\frac{\partial {\sf F}_2}{\partial r_2'}\left(\phi_2^2\right)-
\frac{\partial {\sf F}_3}{\partial r'_3}\left(\phi_3^1\right)=0,\quad
\frac{\partial {\sf F}_1}{\partial z_1'}\left(\phi_1^2\right)+
\frac{\partial {\sf F}_2}{\partial z_2'}\left(\phi_2^2\right)-
\frac{\partial {\sf F}_3}{\partial z'_3}\left(\phi_3^1\right)=0,\label{e17}\\
\frac{\partial {\sf F}_3}{\partial r_3'}\left(\phi_3^2\right)-
\frac{\partial {\sf F}_4}{\partial r_4'}\left(\phi_4^1\right)-
\frac{\partial {\sf F}_5}{\partial r_5'}\left(\phi_5^1\right)=0,\quad
\frac{\partial {\sf F}_3}{\partial z'_3}\left(\phi_3^2\right)-
\frac{\partial {\sf F}_4}{\partial z'_4}\left(\phi_4^1\right)-
\frac{\partial {\sf F}_5}{\partial z'_5}\left(\phi_5^1\right)=0.\nonumber
\eea
Boundary conditions (BC) (\ref{e14}, \ref{e16}) have to be supplemented by
condition of coincidence of interfaces in $C_5,C_6$ located on singular curves
$L_1,L_2$, respectively,
\bea
r_1\left(\phi_1^2\right)=r_2\left(\phi_2^2\right)=r_3\left(\phi_3^1\right),
\hspace{.2cm} r_4\left(\phi_4^1\right)=r_5\left(\phi_5^1\right)=
r_3\left(\phi_3^2\right),\label{e18}\\
z_1\left(\phi_1^2\right)=z_2\left(\phi_2^2\right)=z_3\left(\phi_3^1\right),
\hspace{.2cm} z_4\left(\phi_4^1\right)=z_5\left(\phi_5^1\right)=
z_3\left(\phi_3^2\right),\nonumber
\eea
while the angular coordinates $\phi_j^k$ and $\psi_{\alpha}^j$ are related by
\bea
z_1\left(\phi_1^1\right)\!=\!Z_1\left(\psi_1^1\right),\quad
r_1\left(\phi_1^1\right)\!=\!R_1\left(\psi_1^1\right),\quad
z_2\left(\phi_2^1\right)\!=\!Z_1\left(\psi_1^2\right),\quad
r_2\left(\phi_2^1\right)\!=\!R_1\left(\psi_1^2\right),\label{e19}\\
z_4\left(\phi_4^2\right)\!=\!Z_2\left(\psi_2^4\right),\quad
r_4\left(\phi_4^2\right)\!=\!R_2\left(\psi_2^4\right),\quad
z_5\left(\phi_5^2\right)\!=\!Z_2\left(\psi_2^5\right),\quad
r_5\left(\phi_5^2\right)\!=\!R_2\left(\psi_2^5\right).\nonumber
\eea
Thus, we have 24 BC for the ten YLE (\ref{e13}) of the second order. Let us
arrange them as follows,
\bea
&&\frac{\delta {\sf F}_1}{\delta r_1}=\frac{\delta {\sf F}_1}{\delta z_1}=0,
\quad\left\{\begin{array}{l}
\frac{d R_1}{d\psi_1}\left(\psi_1^1\right)\frac{\partial {\sf F}_1}
{\partial r_1'}\left(\phi_1^1\right)+\frac{d Z_1}{d\psi_1}\left(\psi_1^1\right)
\frac{\partial {\sf F}_1}{\partial z'_1}\left(\phi_1^1\right)+G_1^{me}\left(
\psi_1^1\right)=0,\\
r_1\left(\phi_1^2\right)=r_3\left(\phi_3^1\right),\quad
z_1\left(\phi_1^2\right)=z_3\left(\phi_3^1\right),\\
z_1\left(\phi_1^1\right)\!=\!Z_1\left(\psi_1^1\right),\quad r_1\left(
\phi_1^1\right)\!=\!R_1\left(\psi_1^1\right),\end{array}\right.\hspace{-2cm}
\label{e20}\\
&&\frac{\delta {\sf F}_2}{\delta r_2}=\frac{\delta {\sf F}_2}{\delta z_2}=0,
\quad\left\{\begin{array}{l}
\frac{d R_1}{d\psi_1}\left(\psi_1^2\right)\frac{\partial {\sf F}_2}
{\partial r_2'}\left(\phi_2^1\right)+\frac{d Z_1}{d\psi_1}\left(\psi_1^2\right)
\frac{\partial{\sf F}_2}{\partial z'_2}\left(\phi_2^1\right)+G_1^{im}\left(
\psi_1^2\right)=0,\\
r_2\left(\phi_2^2\right)=r_3\left(\phi_3^1\right),\quad z_2\left(\phi_2^2\right)
=z_3\left(\phi_3^1\right),\\
z_2\left(\phi_2^1\right)\!=\!Z_1\left(\psi_1^2\right),\quad r_2\left(\phi_2^1
\right)\!=\!R_1\left(\psi_1^2\right),\end{array}\right.\label{e21}\\
&&\frac{\delta {\sf F}_3}{\delta r_3}=\frac{\delta {\sf F}_3}{\delta z_3}=0,
\quad\left\{\begin{array}{l}
\frac{\partial {\sf F}_1}{\partial r_1'}\left(\phi_1^2\right)+
\frac{\partial {\sf F}_2}{\partial r_2'}\left(\phi_2^2\right)-
\frac{\partial {\sf F}_3}{\partial r'_3}\left(\phi_3^1\right)=0,\\
\frac{\partial {\sf F}_1}{\partial z_1'}\left(\phi_1^2\right)+
\frac{\partial {\sf F}_2}{\partial z_2'}\left(\phi_2^2\right)-
\frac{\partial {\sf F}_3}{\partial z'_3}\left(\phi_3^1\right)=0,\\
\frac{\partial {\sf F}_3}{\partial r_3'}\left(\phi_3^2\right)-
\frac{\partial {\sf F}_4}{\partial r_4'}\left(\phi_4^1\right)-
\frac{\partial {\sf F}_5}{\partial r_5'}\left(\phi_5^1\right)=0,\\
\frac{\partial {\sf F}_3}{\partial z'_3}\left(\phi_3^2\right)-
\frac{\partial {\sf F}_4}{\partial z'_4}\left(\phi_4^1\right)-
\frac{\partial {\sf F}_5}{\partial z'_5}\left(\phi_5^1\right)=0,\end{array}
\right.\quad\label{e22}\\
&&\frac{\delta {\sf F}_4}{\delta r_4}=\frac{\delta {\sf F}_4}{\delta z_4}=0,
\quad\left\{\begin{array}{l}
\frac{d R_2}{d\psi_2}\left(\psi_2^4\right)\frac{\partial {\sf F}_4}
{\partial r_4'}\left(\phi_4^2\right)+\frac{d Z_2}{d\psi_2}\left(\psi_2^4\right)
\frac{\partial {\sf F}_4}{\partial z'_4}\left(\phi_4^2\right)+G_2^{im}\left(
\psi_2^4\right)=0,\\
r_4\left(\phi_4^1\right)=r_3\left(\phi_3^2\right),\quad r_4\left(
\phi_4^1\right)=r_3\left(\phi_3^2\right),\\
z_4\left(\phi_4^2\right)\!=\!Z_2\left(\psi_2^4\right),\quad r_4\left(\phi_4^2
\right)\!=\!R_2\left(\psi_2^4\right),\end{array}\right.\label{e23}\\
&&\frac{\delta {\sf F}_5}{\delta r_5}=\frac{\delta {\sf F}_5}{\delta z_5}=0,
\quad\left\{\begin{array}{l}\frac{d R_2}{d\psi_2}\left(\psi_2^5\right)
\frac{\partial {\sf F}_5}{\partial r_5'}\left(\phi_5^2\right)+
\frac{d Z_2}{d\psi_2}\left(\psi_2^5\right)\frac{\partial {\sf F}_5}
{\partial z'_5}\left(\phi_5^2\right)+G_2^{me}\left(\psi_2^5\right)=0,\\
r_5\left(\phi_5^1\right)=r_3\left(\phi_3^2\right),\quad z_5\left(\phi_5^1\right)
=z_3\left(\phi_3^2\right),\\
z_5\left(\phi_5^2\right)\!=\!Z_2\left(\psi_2^5\right),\quad r_5\left(\phi_5^2
\right)\!=\!R_2\left(\psi_2^5\right),\end{array}\right.\label{e24}
\eea
where
\bea
&&G_1^{me}\left(\psi_1^1\right)={\sf G}_1^m\left(\psi_1^1\right)-{\sf G}_1^e
\left(\psi_1^1\right),\quad G_1^{im}\left(\psi_1^2\right)={\sf G}_1^i\left(
\psi_1^2\right)-{\sf G}_1^m\left(\psi_1^2\right),\nonumber\\
&&G_2^{im}\left(\psi_2^4\right)={\sf G}_2^i\left(\psi_2^4\right)-{\sf G}_2^m
\left(\psi_2^4\right),\quad G_2^{me}\left(\psi_5^2\right)={\sf G}_2^m
\left(\psi_2^5\right)-{\sf G}_2^e\left(\psi_2^5\right).\nonumber
\eea
\subsection{Curvature law and interface consistency rules}\label{c21}
Analysis of YLE (\ref{e13}) yields an important conclusion about the curvatures 
$H_j$ of five interfaces. Consider (\ref{e13}) and recall that according to 
\cite{FR2014}, $\lambda_j=2\gamma_jH_j$. Combining this scaling with (\ref{e6}) 
we arrive at relationships between the curvatures of three interfaces,
\bea
\gamma_1H_1+\gamma_2H_2=\gamma_3H_3,\quad H_1=H_5,\quad H_2=H_4.\label{e25}
\eea
Simple verification of (\ref{e25}) can be done in special cases. Indeed, if the
liquids {\em i} and {\em m} are indistinguishable, i.e., $\gamma_1=\gamma_3$ and
$\gamma_2=0$, then $H_1=H_3$. On the other hand, if the liquids  {\em m} and
{\em e} are indistinguishable, i.e., $\gamma_2=\gamma_3$ and $\gamma_1=0$, then
$H_2=H_3$. In the case $\gamma_1=\gamma_2=\gamma_3\neq 0$, we arrive at relation
known in theory of double bubble \cite{Morg2009} when three spherical soap
surfaces meet at a contact line.

We can formulate strong statements on consistency of five interfaces based on
relations (\ref{e25}). Recall \cite{FR2014} that there exists only one type,
${\sf Nod^-}$, of interfaces with negative $H$, while the other interfaces have
positive $H$: nodoid ${\sf Nod^+}$, cylinder ${\sf Cyl}$, unduloid ${\sf Und}$,
sphere ${\sf Sph}$, or zero curvature, catenoid ${\sf Cat}$. Denote by ${\sf 
Mns}^+\!\!=\!\!\left\{{\sf Nod^+},{\sf Cyl},{\sf Und},{\sf Sph}\right\}$ a set 
of interfaces with $H>0$ and by ${\sf Mns}^{\pm}\!\!=\!\!\left\{{\sf Mns}^+,
{\sf Cat},{\sf Nod^-}\right\}$ a set of all admissible interfaces. The rules of 
interfaces consistency with curvatures $H_1,H_2,H_3$ are given in Table, e.g., 
if the first and second interfaces are ${\sf Cat}$ and ${\sf Nod^-}$ then the 
third interface has to be also ${\sf Nod^-}$, but if the first and second 
interfaces are ${\sf Und}$ and ${\sf Nod^-}$ then the third interface may be 
any of six interfaces.
\begin{center}
\vspace{-.7cm}
$$
\begin{array}{|c||c|c|c|c|}\hline
{\sf Interfaces} & {\sf Mns}^+ & {\sf Cat} & {\sf Nod^-} \\\hline\hline
{\sf Mns}^+ & {\sf Mns}^+ &{\sf Mns}^+ &{\sf Mns}^{\pm}\\\hline
{\sf Cat}   & {\sf Mns}^+ &{\sf Cat} &{\sf Nod^-} \\\hline
{\sf Nod^-} &{\sf Mns}^{\pm}&{\sf Nod^-}&{\sf Nod^-}\\\hline
\end{array}
$$\label{ta1}
\end{center}
\subsection{Standard parameterization and symmetric setup}\label{c22}
Consider non-zero curvature interfaces $r_j(\phi_j)$, $z_j(\phi_j)$, $1\leq j
\leq 5$, between two solid bodies, $\{R_{\alpha}(\psi_{\alpha})$, $Z_{\alpha}
(\psi_{\alpha})\}$, $\alpha=1,2$, and choose interfaces parameterization in 
such a way that the lower $\phi_j^2$ and the upper $\phi_j^1$ coordinates of 
endpoints $C_1,C_2,C_3,C_4$ are located on the solid surfaces $S_1,S_2$ and 
governed by BC while the other two points $C_5,C_6$ denote the triple points 
located on singular curves $L_1,L_2$ where three different interfaces meet 
together.

Following \cite{FR2014} write the parametric expressions for the shape of such 
interfaces $z_j(\phi_j)$ and $r_j(\phi_j)$,
\bea
z_j(\phi_j)\!=\!\frac{M(\phi_j,B_j)}{2|H_j|}+d_j,\quad
r_j(\phi_j)\!=\!\frac1{2|H_j|}\sqrt{1+B_j^2+2B_j\cos\phi_j},\hspace{.5cm}
\label{e26}\\
M(\phi,B)=(1+B)E\left(\frac{\phi}{2},m\right)+(1-B)F\left(\frac{\phi}{2},m
\right),\quad m^2=\frac{4B}{(1+B)^2},\hspace{.6cm}\nonumber\\
r_j'=-\frac{B_j\sin\phi_j}{2|H_j|r_j},\quad z_j'=\frac{1+B_j\cos\phi_j}{2|H_j|
r_j},\quad\frac{z_j'}{r_j'}=-\frac{1+B_j\cos\phi_j}{B_j\sin\phi_j},\quad
r_j'^2+z_j'^2=1.\hspace{.5cm}\label{e27}
\eea
For all interfaces we have to find 24 unknowns: 15-1=14 interfaces parameters
$H_j,B_j,d_j$ (due to (\ref{e25})) and 10 endpoint values $\phi_j^1,\phi_j^2$.
These unknowns have to satisfy 24 BCs in (\ref{e20}-\ref{e24}).

When both surfaces $S_1$ and $S_2$ are similar and the picture in Figure
\ref{f1}b is symmetric w.r.t. midline between $S_1$ and $S_2$, then such setup
reduces the problem above to six YLE (\ref{e20}-\ref{e22}) for the first, 
second and third interfaces with twelve unknowns:
\bea
\phi_1^1,\;\phi_2^1,\;\phi_3^1,\;\phi_1^2,\;\phi_2^2,\;d_1,\;d_2,\;B_1,\;B_2,
\;B_3,\;\;H_1,\;H_2,\nonumber
\eea
and $\phi_3^2=\pi,\;2d_3\!=\!-M(\pi,B_3)/|H_3|$ and $H_3=(\gamma_1H_1+\gamma_2
H_2)/\gamma_3$. This number coincides with twelve BCs which comprise ten BCs in 
(\ref{e21},\ref{e22}) and two first BCs in (\ref{e23}). Calculate the partial 
derivatives $\partial {\sf F}_j/\partial r_j'$, $\partial {\sf F}_j/\partial 
z_j'$ and write these twelve BCs,
\bea
&&\gamma_1r_1\left(r_1'R_1'+z_1'Z_1'\right)+\left(\gamma_{s_1}^m-\gamma_{s_1}^e
\right)R_1\sqrt{R_1'^2+Z_1'^2}+\lambda_1Z_1'\frac{R_1^2-r_1^2}{2}=0,\quad
\phi_1=\phi_1^1,\;\psi_1=\psi_1^1,\quad\nonumber\\
&&r_1\left(\phi_1^2\right)=r_3\left(\phi_3^1\right),\quad z_1\left(\phi_1^2
\right)=z_3\left(\phi_3^1\right),\quad z_1\left(\phi_1^1\right)\!=\!Z_1\left(
\psi_1^1\right),\quad r_1\left(\phi_1^1\right)\!=\!R_1\left(\psi_1^1\right),
\hspace{1cm}\label{e29}\\
&&\gamma_2r_2\left(r_2'R_1'+z_2'Z_1'\right)+\left(\gamma_{s_1}^i-\gamma_{s_1}^m
\right)R_1\sqrt{R_1'^2+Z_1'^2}+\lambda_2Z_1'\frac{R_1^2-r_2^2}{2}=0,\quad
\phi_2=\phi_2^1,\;\psi_1=\psi_1^2,\nonumber\\
&&r_2\left(\phi_2^2\right)=r_3\left(\phi_3^1\right),\quad z_2\left(\phi_2^2
\right)=z_3\left(\phi_3^1\right),\quad z_2\left(\phi_2^1\right)\!=\!Z_1\left(
\psi_1^2\right),\quad r_2\left(\phi_2^1\right)\!=\!R_1\left(\psi_1^2\right),
\label{e30}\\
&&\gamma_1r_1r_1'+\gamma_2r_2r_2'-\gamma_3r_3r_3'=0,\hspace{.5cm}
\phi_1=\phi_1^2,\;\phi_2=\phi_2^2,\;\phi_3=\phi_3^1,\nonumber\\
&&\gamma_1r_1z_1'+\gamma_2r_2z_2'-\gamma_3r_3z_3'=\frac1{2}
\left(\lambda_1r_1^2+\lambda_2r_2^2-\lambda_3r_3^2\right).\label{e31}
\eea
After simplification we obtain
\bea
&&\gamma_1\cos\theta_1^1+\gamma_{s_1}^m-\gamma_{s_1}^e=0,\quad\gamma_2\cos\theta
_1^2+\gamma_{s_1}^i-\gamma_{s_1}^m=0,\label{e32}\\
&&r_1\left(\phi_1^2\right)=r_2\left(\phi_2^2\right)=r_3\left(\phi_3^1\right),
\quad z_1\left(\phi_1^1\right)\!=\!Z_1\left(\psi_1^1\right),\quad r_1\left(
\phi_1^1\right)\!=\!R_1\left(\psi_1^1\right),\nonumber\\
&& z_1\left(\phi_1^2\right)=z_2\left(\phi_2^2\right)=z_3\left(\phi_3^1\right),
\quad z_2\left(\phi_2^1\right)\!=\!Z_1\left(\psi_1^2\right),\quad
r_2\left(\phi_2^1\right)\!=\!R_1\left(\psi_1^2\right),\nonumber\\
&&\gamma_1r_1'\left(\phi_1^2\right)+\gamma_2r_2'\left(\phi_2^2\right)-\gamma_3
r_3'\left(\phi_3^1\right)=0,\quad \gamma_1z_1'\left(\phi_1^2\right)+\gamma_2z_2'
\left(\phi_2^2\right)-\gamma_3z_3'\left(\phi_3^1\right)=0,\hspace{1cm}
\label{e33}
\eea
where $\;\cos\theta_1^j\!=\!\left(r_j'R_1'+z_j'Z_1'\right)\!/\!\sqrt{R_1'^2+
Z_1'^2}$ determines the contact angle $\theta_1^j$ of the $j$-th interface and
$S_1$. Two equalities in (\ref{e32}) give the Young relations at the points
$C_1,C_2$ on $S_1$  while two equalities in (\ref{e33}) represent the vectorial 
Young relation at the triple point $C_5$ located on a singular curve. Indeed, 
the latter equalities are the $r$ and $z$ projections of the vectorial equality 
for capillary forces ${\bf f}_j$ at $C_5$ in outward directions w.r.t. $C_5$
and tangential to meridional section of menisci,
\bea
{\bf f}_1(C_5)+{\bf f}_2(C_5)+{\bf f}_3(C_5)=0,\quad {\bf f}_j(C_5)=\gamma_j
\left\{r_j'(C_5),z_j'(C_5)\right\}.\label{e34}
\eea
Finish this section with one more observation related the surface tensions
$\gamma_j$ and contact angles of three interfaces on solid surface. Bearing in
mind that $\gamma_3\cos\theta_1^3+\gamma_{s_1}^i-\gamma_{s_1}^e=0$, combine the
last equality with two others in (\ref{e32}) and obtain,
\bea
\gamma_1\cos\theta_1^1+\gamma_2\cos\theta_1^2=\gamma_3\cos\theta_1^3.\label{e35}
\eea
\subsection{Solving the BC equations (liquid bridges between two parallel 
plates)}\label{c23}
Making use of standard parametrization (\ref{e26}) we present below twelve BCs
(\ref{e32},\ref{e33}) for twelve unknowns $\phi_1^1,\phi_2^1,\phi_3^1,\phi_1^2,
\phi_2^2,d_1,d_2,B_1,B_2,B_3,H_1,H_2,$ in a way convenient for numerical 
calculations,
\bea
&&\frac{\gamma_1B_1\sin\phi_1^2}{|H_1|}+\frac{\gamma_2B_2\sin\phi_2^2}{|H_2|}
=\frac{\gamma_3B_3\sin\phi_3^1}{|H_3|},\quad
\frac{\gamma_1B_1\cos\phi_1^2}{|H_1|}+\frac{\gamma_2B_2\cos\phi_2^2}{|H_2|}
=\frac{\gamma_3B_3\cos\phi_3^1}{|H_3|},\nonumber\\
&&\frac{\sqrt{1+2B_1\cos\phi_1^2+B_1^2}}{|H_1|}=\frac{\sqrt{1+2B_2\cos\phi_2^2+
B_2^2}}{|H_2|}=\frac{\sqrt{1+2B_3\cos\phi_3^1+B_3^2}}{|H_3|},\hspace{2.7cm}
\label{q1}\\
&&\frac{M(\phi_1^2,B_1)}{2|H_1|}+d_1=\frac{M(\phi_2^2,B_2)}{2|H_2|}+d_2=\frac{
M(\phi_3^1,B_3)}{2|H_3|}+d_3,\quad d_3=-\frac{M(\pi,B_3)}{2|H_3|},\nonumber\\
&&\frac{M(\phi_j^1,B_j)-M(\phi_j^2,B_j)}{2|H_j|}=Z_1\left(\psi_1^j\right)-
\frac{M(\phi_3^1,B_3)}{2|H_3|}-d_3,\quad\nonumber\\
&&|H_j|=\frac{\sqrt{1+2B_j\cos\phi_j^1+B_j^2}}{2R_1\left(\psi_1^j\right)},\quad
B_j=\left[\cos\phi_j^1+\sin\phi_j^1\tan\theta_1^j\right]^{-1},\quad j=1,2,
\nonumber
\eea
where $H_3=H_1\gamma_1/\gamma_3+H_2\gamma_2/\gamma_3$.

The numerical optimization of the solution was done by a standard gradient 
descent algorithm. The cost function for the optimization problem was chosen to 
be the weighted sum of absolute values of the differences between the right and 
the left hand sides of the six first equations in (\ref{q1}).
\begin{figure}[h!]\begin{center}\begin{tabular}{ccc}
\psfig{figure=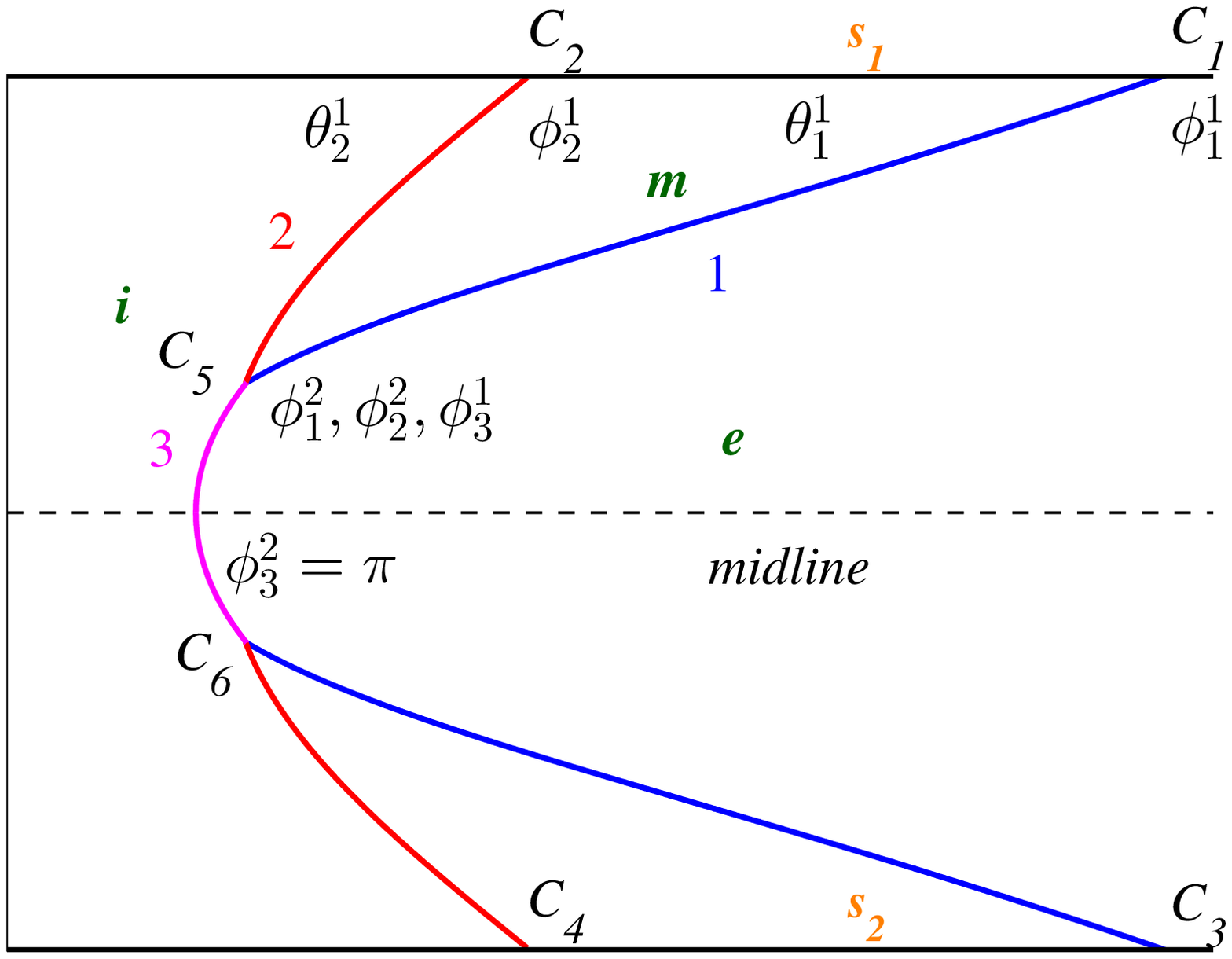,height=6.5cm}
\hspace{.5cm}&\hspace{.5cm}
\psfig{figure=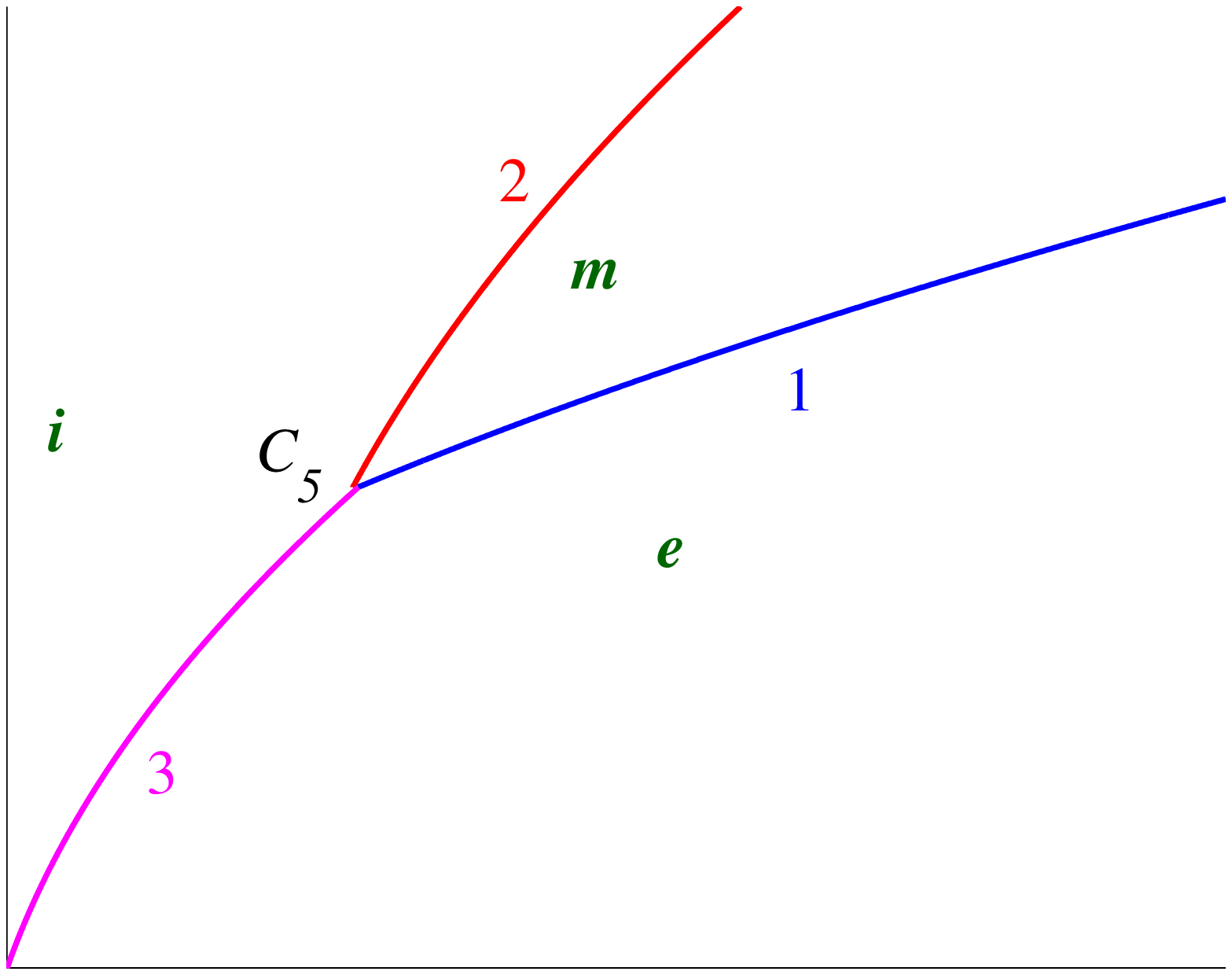,height=6cm,width=6cm}\\
(a)&(b)
\end{tabular}\end{center}
\vspace{-.2cm}
\caption{a) Five ${\sf Und}$ interfaces for three immiscible media ($i$ -- 
water, $m$ -- octane, $e$ -- air) trapped between two similar solid plates 
($Z_1-Z_2=2$) with free BC. b) Enlarged view of the vicinity of the triple 
point $C_5$ on the singular curve where three phases coexist. The angles 
between the adjacent interfaces read: $\Phi_{12}=36.76^o$, $\Phi_{23}=159.43^o$,
$\Phi_{31}=163.81^o$.}\label{f3}
\end{figure}

In Figure \ref{f3} we present the shapes of five interfaces of different 
curvatures for three immiscible media: $i$ -- water, $m$ -- octane ($C_8H_{18}$,
a component of petrol), $e$ -- air, trapped between two similar glass plates 
with free BCs and capillary parameters taken from \cite{ttt}. The 
interfaces have the following parameters,
\begin{center}
$$
\begin{array}{|c||c|c|c|c|c|c|c|}\hline
{\sf Interfaces} & \gamma_j,\;$mN/m$&\theta_j& B_j & H_j & d_j&\phi_j^2&
\phi_j^1\\ \hline\hline
1 (e-m)& 21.8 & 19^o   & 0.959 & 0.095 & -11.424 & 184.39^o & 208.74^o\\\hline
2 (m-i)& 50.8 & 39^o   & 0.778 & 0.272 & -4.664 & 186.405^o & 216.34^o\\\hline
3 (e-i)& 72.8 & 34.4^o & 0.855 & 0.218 & -5.599 & 180^o     & 188.14^o\\\hline
\end{array}
$$\label{ta2}
\end{center}

\noindent
The volumes of liquids confined inside interfaces read $V_m=4.009$, $V_i=2.674$.
\section{Variational problem for three interfaces}\label{c3}
Consider a functional $E[r,z]$ of surface energy
\bea
E[r,z]\!=\!\sum_{j=1}^3\int_{\phi_j^2}^{\phi_j^1}\!\!{\sf E}_jd\phi_j+\!
\int_0^{\psi_1^2}\!\!{\sf A}_{s_1}^id\psi_1+\!\int_{\psi_1^2}^{\psi_1^1}\!\!
{\sf A}_{s_1}^md\psi_1+\!\int_{\psi_1^1}^{\infty}\!\!{\sf A}_{s_1}^ed\psi_1+\!
\int_0^{\psi_2^3}{\sf A}_{s_2}^i\psi_2+\!\int_{\psi_2^3}^{\infty}\!\!{\sf A}_
{s_2}^e\psi_2,\label{k1}
\eea
and two functionals $V_i[r,z]$ and $V_m[r,z]$ of volumes of the $i$ and $m$
liquids
\bea
&&V_m[r,z]=\int_{\phi_1^2}^{\phi_1^1}\!{\sf V}_1d\phi_1-\int_{\phi_2^2}^{\phi
_2^1}\!{\sf V}_2d\phi_2-\int_{\psi_1^2}^{\psi_1^1}{\sf B}_{s_1}d\psi_1,
\label{k2}\\
&&V_i[r,z]=\int_{\phi_2^2}^{\phi_2^1}{\sf V}_2d\phi_2+\int_{\phi_3^2}^{\phi_3^1}
{\sf V}_3d\phi_3-\int_0^{\psi_1^2}{\sf B}_{s_1}d\psi_1+\int_0^{\psi_2^3}
{\sf B}_{s_2}d\psi_2,\nonumber
\eea
where all integrands ${\sf E}_j$, ${\sf A}_{s_{\alpha}}^{i,m,e}$, ${\sf V}_j$
and ${\sf B}_{s_{\alpha}}$ are defined in (\ref{e2}, \ref{e4}).

Consider the composed functional $W[r,z]=E[r,z]-\lambda_1V_m[r,z]-\lambda_3V_i
[r,z]$ and represent it in the following form,
\bea
W[r,z]=\sum_{j=1}^3\!\int_{\phi_j^2}^{\phi_j^1}\!\!{\sf F}_jd\phi_1\!+\!\int_0^
{\psi_1^2}\!\!{\sf G}_1^id\psi_1+\int_{\psi_1^2}^{\psi_1^1}\!\!{\sf G}_1^md\psi_
1\!+\!\int_{\psi_1^1}^{\infty}\!\!{\sf G}_1^ed\psi_1\!-\!\int_0^{\psi_2^3}\!\!
{\sf G}_2^id\psi_2\!-\!\int_{\psi_2^3}^{\infty}\!\!{\sf G}_2^ed\psi_2,\label{k3}
\eea
where the integrands are given in (\ref{e6}).

Applying a similar technique as in section \ref{c2} we arrive at the first
variation,
\bea
\delta
W&=&\sum_{j=1}^5\left[\int_{\phi_j^2}^{\phi_j^1}\left(u_j\frac{\delta{\sf F}_j}
{\delta r_j}+v_j\frac{\delta {\sf F}_j}{\delta z_j}\right)d\phi_j+\left(u_j
\frac{\partial {\sf F}_j}{\partial r_j'}+v_j\frac{\partial {\sf F}_j}{\partial
z'_j}\right)_{\phi_j^2}^{\phi_j^1}\right]+\label{k4}\\
&&\left({\sf G}_1^i-{\sf G}_1^m\right)\delta\psi_1^2+\left({\sf G}_1^m-{\sf G}_
1^e\right)\delta\psi_1^1-\left({\sf G}_2^i-{\sf G}_2^e\right)\delta\psi_2^3.
\nonumber
\eea

This case does not allow a symmetric version and therefore is less reducible
compared to the 5 interface case w.r.t. the number of unknowns and BC equations.
This number equal fifteen: nine interface parameters $H_j,B_j,d_j,$ and six
endpoint values $\phi_j^1,\phi_j^2$. They satisfy fifteen BC equations
\bea
&&\frac{\delta {\sf F}_1}{\delta r_1}=\frac{\delta {\sf F}_1}{\delta z_1}=0,
\quad\left\{\begin{array}{l}
\frac{d R_1}{d\psi_1}\left(\psi_1^1\right)\frac{\partial {\sf F}_1}
{\partial r_1'}\left(\phi_1^1\right)+\frac{d Z_1}{d\psi_1}\left(\psi_1^1\right)
\frac{\partial {\sf F}_1}{\partial z'_1}\left(\phi_1^1\right)+G_1^{me}\left(
\psi_1^1\right)=0,\\
r_1\left(\phi_1^2\right)=r_3\left(\phi_3^1\right),\quad
z_1\left(\phi_1^2\right)=z_3\left(\phi_3^1\right),\\
z_1\left(\phi_1^1\right)\!=\!Z_1\left(\psi_1^1\right),\quad r_1\left(\phi_1^1
\right)\!=\!R_1\left(\psi_1^1\right),\end{array}\right.\hspace{-2cm}
\nonumber\\
&&\frac{\delta {\sf F}_2}{\delta r_2}=\frac{\delta {\sf F}_2}{\delta z_2}=0,
\quad\left\{\begin{array}{l}
\frac{d R_1}{d\psi_1}\left(\psi_1^2\right)\frac{\partial {\sf F}_2}{\partial
r_2'}\left(\phi_2^1\right)+\frac{d Z_1}{d\psi_1}\left(\psi_1^2\right)\frac{
\partial{\sf F}_2}{\partial z'_2}\left(\phi_2^1\right)+G_1^{im}\left(\psi_1^2
\right)=0,\\
r_2\left(\phi_2^2\right)=r_3\left(\phi_3^1\right),\quad z_2\left(\phi_2^2\right)
=z_3\left(\phi_3^1\right),\\
z_2\left(\phi_2^1\right)\!=\!Z_1\left(\psi_1^2\right),\quad r_2\left(\phi_2^1
\right)\!=\!R_1\left(\psi_1^2\right),\end{array}\right.\label{k5}\\
&&\frac{\delta {\sf F}_3}{\delta r_3}=\frac{\delta {\sf F}_3}{\delta z_3}=0,
\quad\left\{\begin{array}{l}
\frac{\partial {\sf F}_1}{\partial r_1'}\left(\phi_1^2\right)+
\frac{\partial {\sf F}_2}{\partial r_2'}\left(\phi_2^2\right)-
\frac{\partial {\sf F}_3}{\partial r'_3}\left(\phi_3^1\right)=0,\\
\frac{\partial {\sf F}_1}{\partial z_1'}\left(\phi_1^2\right)+
\frac{\partial {\sf F}_2}{\partial z_2'}\left(\phi_2^2\right)-
\frac{\partial {\sf F}_3}{\partial z'_3}\left(\phi_3^1\right)=0,\\
\frac{d R_2}{d\psi_2}\left(\psi_2^3\right)\frac{\partial {\sf F}_5}
{\partial r_5'}\left(\phi_3^2\right)+\frac{d Z_2}{d\psi_2}\left(\psi_2^3\right)
\frac{\partial {\sf F}_5}{\partial z'_5}\left(\phi_3^2\right)+G_2^{ie}\left(
\psi_2^3\right)=0,\\
z_5\left(\phi_5^2\right)\!=\!Z_2\left(\psi_2^3\right),\quad r_5\left(\phi_5^2
\right)\!=\!R_2\left(\psi_2^3\right).\end{array}\right.\nonumber
\eea
that gives
\bea
&&\gamma_1\cos\theta_1^1+\gamma_{s_1}^m-\gamma_{s_1}^e=0,\quad\gamma_2\cos\theta
_1^2+\gamma_{s_1}^i-\gamma_{s_1}^m=0,\quad\gamma_3\cos\theta_2^3+\gamma_{s_2}^i
-\gamma_{s_2}^e=0,\label{k6}\\
&&\gamma_1r_1'\left(\phi_1^2\right)+\gamma_2r_2'\left(\phi_2^2\right)-\gamma_3
r_3'\left(\phi_3^1\right)=0,\quad\gamma_1z_1'\left(\phi_1^2\right)+\gamma_2z_2'
\left(\phi_2^2\right)-\gamma_3z_3'\left(\phi_3^1\right)=0,\nonumber\\
&&r_1\left(\phi_1^2\right)=r_2\left(\phi_2^2\right)=r_3\left(\phi_3^1\right),
\quad z_1\left(\phi_1^2\right)=z_2\left(\phi_2^2\right)=z_3\left(\phi_3^1
\right),\nonumber\\
&&r_1\left(\phi_1^1\right)\!=\!R_1\left(\psi_1^1\right),\quad
r_2\left(\phi_2^1\right)\!=\!R_1\left(\psi_1^2\right),\quad
r_3\left(\phi_3^2\right)=R_2\left(\psi_2^3\right),\label{k8}\\
&&z_1\left(\phi_1^1\right)\!=\!Z_1\left(\psi_1^1\right),\quad
z_2\left(\phi_2^1\right)\!=\!Z_1\left(\psi_1^2\right),\quad
z_3\left(\phi_3^2\right)=Z_2\left(\psi_2^3\right).\nonumber
\eea
Three equalities in (\ref{k6}) cannot be reduced to a single equality similar
to (\ref{e35}) if the upper and lower solid bodies have different capillary
properties, namely, $\gamma_{s_2}^i-\gamma_{s_1}^i\neq \gamma_{s_2}^e-
\gamma_{s_1}^e$, i.e.,
\bea
\gamma_1\cos\theta_1^1+\gamma_2\cos\theta_1^2\neq\gamma_3\cos\theta_2^3.
\nonumber
\eea
\subsection{Solving the BC equations (liquid bridges between two parallel
plates)}\label{c31}
Using a standard parametrization (\ref{e26}) and relation (\ref{e25}) for $H_3$
we present below fourteen BCs (\ref{k6},\ref{k8}) for fourteen unknowns:
$\phi_1^1,\phi_2^1,\phi_3^1,\phi_1^2,\phi_2^2,\phi_3^2,d_1,d_2,d_3,B_1,B_2,B_3,
H_1,H_2$ in a way convenient for numerical calculations,
\bea
&&B_j=[\cos\phi_j^1+\sin\phi_j^1\tan\theta_1^j]^{-1},\quad\frac{M(\phi_j^1,B_j)}
{2|H_j|}+d_j=Z_1\left(\psi_1^j\right),\quad j=1,2,\nonumber\\
&&B_3=[\cos\phi_3^2+\sin\phi_3^2\tan\theta_2^3]^{-1},
\quad\frac{M(\phi_3^2,B_3)}{2|H_3|}+d_3=Z_2\left(\psi_2^3\right),\nonumber\\
&&|H_1|=\frac{\sqrt{1+2B_1\cos\phi_1^1+B_1^2}}{2R_1\left(\psi_1^1\right)},\quad
|H_2|=\frac{\sqrt{1+2B_2\cos\phi_2^1+B_2^2}}{2R_1\left(\psi_1^2\right)},
\label{k9}\\
&&\frac{\gamma_1B_1\sin\phi_1^2}{|H_1|}+\frac{\gamma_2B_2\sin\phi_2^2}{|H_2|}
=\frac{\gamma_3B_3\sin\phi_3^1}{|H_3|},\quad
\frac{\gamma_1B_1\cos\phi_1^2}{|H_1|}+\frac{\gamma_2B_2\cos\phi_2^2}{|H_2|}
=\frac{\gamma_3B_3\cos\phi_3^1}{|H_3|},\nonumber\\
&&\frac{\sqrt{1+2B_1\cos\phi_1^2+B_1^2}}{|H_1|}=\frac{\sqrt{1+2B_2\cos\phi_2^2+
B_2^2}}{|H_2|}=\frac{\sqrt{1+2B_3\cos\phi_3^1+B_3^2}}{|H_3|},\nonumber\\
&&\frac{M(\phi_1^2,B_1)}{2|H_1|}+d_1=\frac{M(\phi_2^2,B_2)}{2|H_2|}+d_2=
\frac{M(\phi_3^1,B_3)}{2|H_3|}+d_3,\nonumber
\eea
where $H_3=(H_1\gamma_1+H_2\gamma_2)/\gamma_3$.
\begin{figure}[h!]\begin{center}\begin{tabular}{ccc}
\psfig{figure=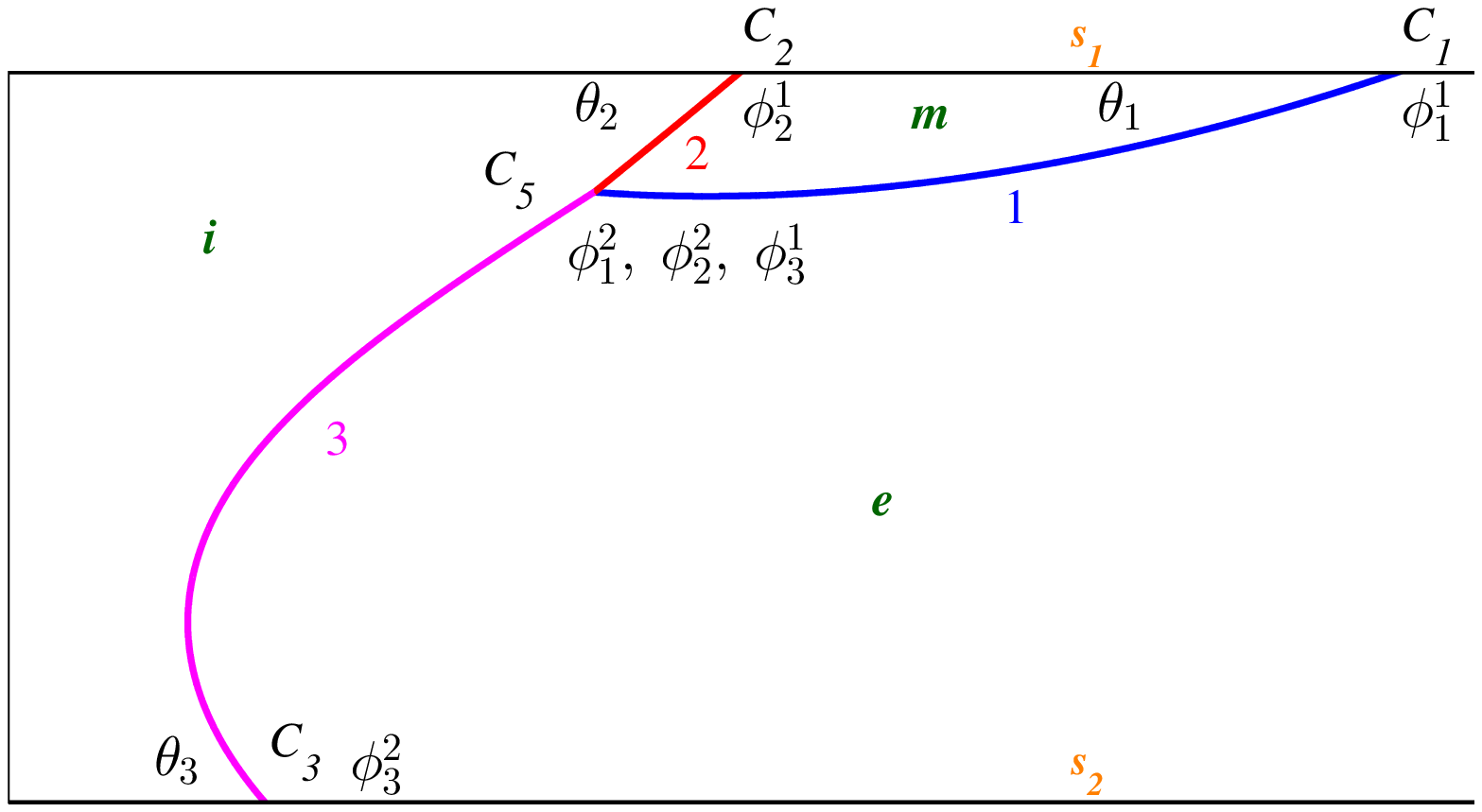,height=6cm,width=9.2cm}&
\psfig{figure=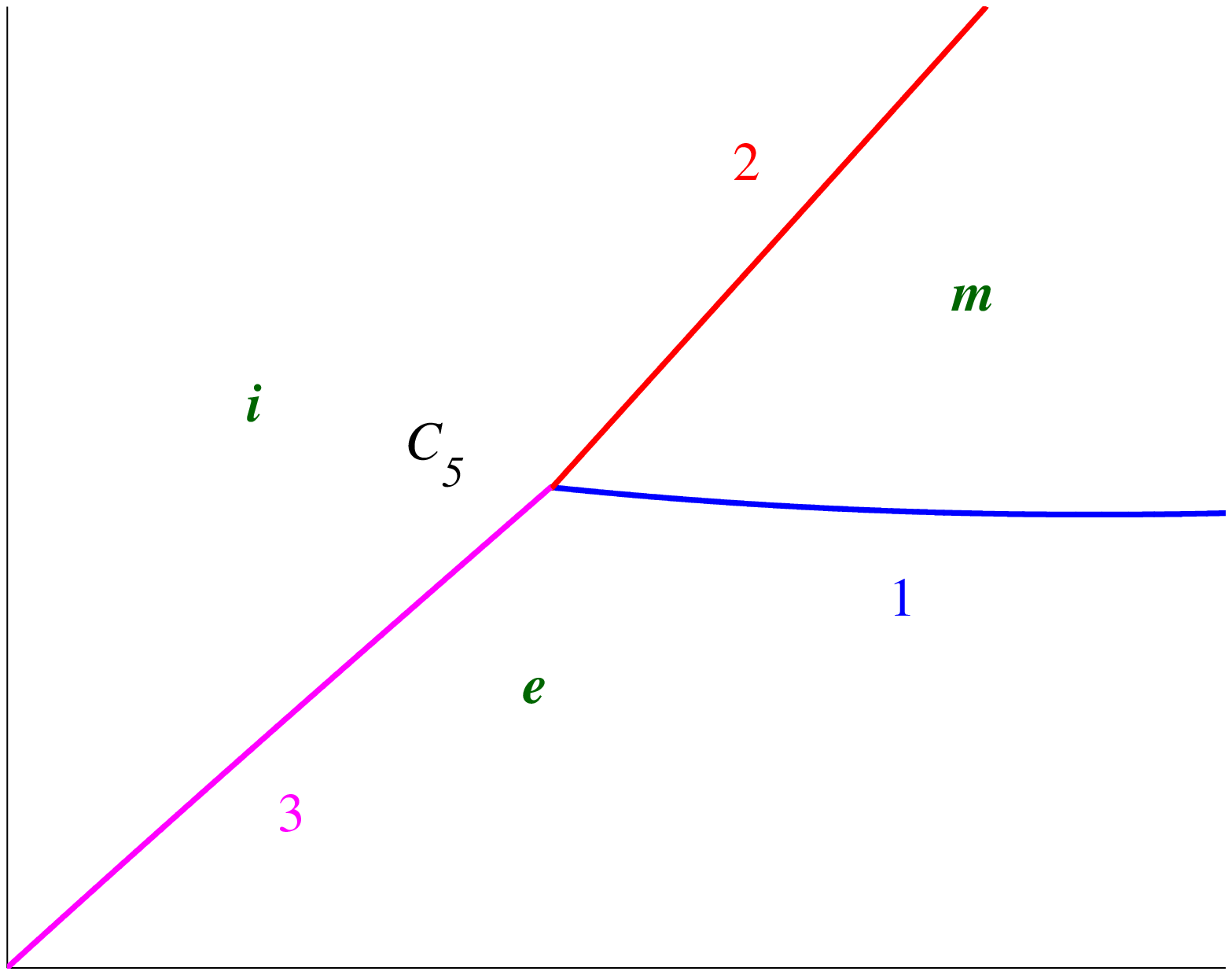,height=5.5cm,width=5.5cm}\\
(a)&(b)
\end{tabular}\end{center}
\vspace{-.2cm}
\caption{a) Two ${\sf Und}$ and one ${\sf Nod}$ interfaces for three immiscible 
media ($i$ -- water, $m$ -- hexane, $e$ -- air) trapped between two (not 
similar) solid plates ($Z_1-Z_2=1$) with free BC. b) Enlarged view of the 
vicinity of the triple point $C_5$ on singular curve $L$ where three phases 
coexist. The angles between the adjacent interfaces reach the following values: 
$\Phi_{12}=46.04^o$, $\Phi_{23}=173.28^o$, $\Phi_{31}=140.69^o$.}\label{f4}
\end{figure}
In Figure \ref{f4} we present the shapes of three interfaces of different
curvatures for three immiscible media: $i$ -- water, $m$ -- hexane ($C_6H_{14}$,
a component of petrol), $e$ -- air, trapped between two plates composed of 
different materials (glass and glass coated with polymer film) with free BCs and
capillary parameters taken from \cite{ttt}. The interfaces have the following 
parameters,
\begin{center}
$$
\begin{array}{|c||c|c|c|c|c|c|c|}\hline
{\sf Interfaces} & \gamma_j\;$mN/m$&\theta_j& B_j & H_j&d_j&\phi_j^2&\phi_j^1\\
\hline\hline
1 (e-m)& 18.4 & 19^o & 1.091 & 0.229 & -2.521 & 199.51^o & 228.89^o\\\hline
2 (m-i)& 51.1 & 40^o & 0.775 & 0.379 & -3.257 & 217.19^o & 228.50^o\\\hline
3 (e-i)& 72.8 & 49^o & 0.841 & 0.324 & -3.574 & 169.79^o & 211.11^o\\\hline
\end{array}
$$\label{ta3}
\end{center}

\noindent
The volumes of liquids confined inside interfaces read $V_m=0.4377$, 
$V_i=0.8940$.
\section{Conclusion}\label{c4}
We formulate a variational problem for coexistence of axisymmetric interfaces of
three immiscible liquids: two of them, {\em i} and {\em m}, immersed in a 
third liquid (or gas) {\em e} and trapped between two smooth solid bodies with 
axisymmetric surfaces $S_1,S_2$ and free contact lines. Assuming the volume
constraints of two liquids {\em i} and {\em m}, we find the governing 
(Young-Laplace) equations (\ref{e13}) supplemented by boundary conditions and 
Young relation (\ref{e14}) on $S_1,S_2$ and transversality relations (\ref{e17})
on singular curve where all liquids meet together. 

We consider two different cases when the problem allows the coexistence of five 
(section \ref{c2}) or three (section \ref{c3}) interfaces. In the first case 
the problem is reduced solving 16 boundary conditions, 4 Young relations and 4 
transversality relations (\ref{e20}-\ref{e24}), i.e., 24 equations for 24 
variables. In the second case this number is reduced substantially, namely, 15 
equations with 15 variables (\ref{k5}) including 10 boundary conditions, 3 
Young relations and 2 transversality relations.

We derive the relationship (\ref{e25}) combining the constant mean curvatures 
of three different interfaces, $e-m$, $m-i$, $e-i$, and give consistency rules 
for interface coexistence (section \ref{c21}).

Another result is the vectorial Young relation (\ref{e34}) at the triple point
which is located on a singular curve. It has a clear physical interpretation as 
the balance equation of capillary forces. More importantly, it gives a new 
insight on an old assertion about the usual Young relations (\ref{e32},\ref{k6})
at a solid/liquid/gas interface refered by R. Finn \cite{Fin2006} to T. Young: 
{\em the contact angle at a solid/liquid/gas interface is a physical constant 
depending only on the materials, and in no other way on the particular 
conditions of problem}, and a well known contradiction with uncompensated normal
force reaction of solid (see \cite{Fin2006} and references therein). Indeed, 
being applied to the contact line of three continuous media, liquid-gas-solid, 
vectorial relation (\ref{e34}) assumes a singular deformation of solid surface 
if its elastic modules take finite values.
\section*{Acknowledgement}
The research was supported by the Kamea Fellowship.


\end{document}